\documentclass[conference]{IEEEtran}
\usepackage{listings}

  \usepackage[pdftex]{graphicx}
  \graphicspath{{../pdf/}{../jpeg/}}
   \DeclareGraphicsExtensions{.pdf,.jpeg,.png}

\usepackage{algorithmic}
\newsavebox{\ieeealgbox}

\usepackage{multicol}
\usepackage{array,multirow}
\usepackage{rotating}

\begin{document}
%
\title{Benchmarking, System Design and Case-studies for Multi-core based Embedded Automotive Systems}

\author{
\begin{tabular}[t]{c@{\extracolsep{1em}}c}
Piotr Dziurzanski, Amit Kumar Singh, Leandro S. Indrusiak & Bj\"{o}rn Saballus\\
Department of Computer Science & Robert Bosch GmbH\\
University of York  & Corporate Sector Research and Advance \\
Deramore Lane, Heslington, York, YO10 5GH, UK & Engineering - Software  (CR/AEA2)\\
 & Postfach 30 02 40, 70442 Stuttgart, Germany\\
\{PD678, Amit.Singh, Leandro.Indrusiak\}@york.ac.uk & bjoern.saballus@de.bosch.com
\end{tabular}
}


\maketitle

\begin{abstract}
In this paper, using of automotive use cases as benchmarks for real-time system design has been proposed. The use cases are described in a format supported by AMALTHEA platform, which is a model based open source development environment for automotive multi-core systems. An example of a simple Electronic Control Unit has been analysed and presented with enough details to reconstruct this system in any format. For researchers willing to use AMALTHEA file format directly, an appropriate parser has been developed and offered\footnote{\label{DreamCloudWebPage}http://www.dreamcloud-project.org
}. An example of applying this parser and benchmark for optimising makespan while not violating the timing constraints by allocating functionality to different Network on Chip resource is demonstrated.
\end{abstract}


%
\IEEEpeerreviewmaketitle

\section{Introduction}

Evaluating automotive real-time multicore system design algorithms using some industrial application benchmarks can be viewed as a factor to increase the reliability of the algorithms. Since a limited number of these benchmarks is publicly available, new sets of industrial use cases are usually welcomed by the real-time research community \cite{Qadri10}. One of such benchmarks is an autonomous vehicle, introduced in \cite{Shi10}, another a real-time jet engine performance calculator \cite{Qadri10}. Some researchers address the lack of application benchmarks by generating synthetic benchmarks, imitating real industrial processes  \cite{Burkimsher14}.

According to \cite{AutomotiveReport14}, one of the most crucial issues of automotive industry is to boost the technology to reduce emissions and increase fuel economy of cars. Further, in the same report it is being forecasted that the market for automotive engine management systems will grow 8.09\% annually, reaching 197 billion USD by 2019. Therefore, sooner or later,  the growing importance of automotive system benchmarking can be predicted.

Some basic automotive algorithms, including controller area network (CAN), tooth-to-spark, angle-to-time conversion, road speed calculation, and table lookup and interpolation are present in AUTOBENCH\textsuperscript{TM} suite from the Embedded Microprocessor Benchmark Consortium (EEMBC). The same consortium cooperates with Volkswagen Group to establish a benchmark suite for microcontrollers aimed at making automotive end products more energy efficient and robust. These benchmark suites are licensed for corporate or academia researchers and thus their availability is limited.

MiBench \cite{Guthaus01} is a set of 35 embedded applications for benchmarking purposes. Six C-codes are categorised as automotive and industrial control, but they are just standard algorithms (basic math, bitcount, quick sort, and image recognition) used often in that domain. However, no whole automotive application implemented in an Electronic Control Units (ECUs) can be found in this suite. For automotive multi-core microcontroller, more realistic use cases, such as described in \cite{Park14}, are needed.

Recently, a model based open source development environment for automotive multi-core systems called AMALTHEA has been developed \cite{AMALTHEA15}. The resulting tool platform, distributed under an Eclipse public license, supports multi-core automotive systems compatible with the AUTOSAR (AUTomotive Open System ARchitecture) standard \cite{AUTOSAR15}. In this paper, a possibility of using the AMALTHEA platform for generating a new suites of automotive benchmarks is investigated. DemoCar, a simple engine control application provided as an example in the AMALTHEA platform distribution is presented and analysed. Since the authors of the AMALTHEA environment has not provided a C-based code for parsing AMALTHEA file, we have developed a simple C++ parser and offered it to the community\textsuperscript{\ref{DreamCloudWebPage}}. The architecture and main functions of this tool are explained in this paper. As a case-study, one example of a system design using this ECU is demonstrated.

In the following section, some basic information about ECUs are provided, including an AMALTHEA automotive application model description.  Then, in section \ref{sec:FileFormat}, AMALTHEA file format is explained. Section \ref{sec:DemoCar} provides details about the DemoCar ECU, which can be parsed by an open-source parser\textsuperscript{\ref{DreamCloudWebPage}} prepared by the authors to read AMALTHEA format files and presented in section \ref{sec:Parser}.  An example of its usage in a genetic-algorithm-based resource allocation is presented in section \ref{sec:CaseStudy}. Section \ref{sec:Conclusions} concludes the paper.

\section{Background information}\label{sec:Background}

Electronic Control Units (ECU) are omnipresent in contemporary cars. Their primary role is to measure, control and steer physical actuators of cars like braking, engine control, transmission systems, head units, car multimedia systems, etc. There exists a number of ECU standards. Component based software design model is standardized by AUTOSAR \cite{AUTOSAR15}, which introduces four basic elements of abstract representation of an automotive application:

\begin{itemize}
\item executable entities (called runnables),
\item events that trigger execution of runnables,
\item labels representing data elements that can be read or written by runnables,
\item data dependencies between runnables and labels.
\end{itemize}

A runnable can be seen as an atomic execution entity, generated from, e.g., a function written in C language or a Simulink model. Each runnable is triggered with an event. After its occurrence, the runnable should be executed and finished before its deadline. Events may be either periodic (for instance, occurring every 5ms) or aperiodic (for instance, depending on value of certain labels). Runnables are executed in order respecting the dependencies between read and written labels so that writers are to be executed prior to readers of the same label. If more than one event occurs at the same instant, they can be handled in a nondeterministic order.

An AUTOSAR model, however, is too detailed for performing fast design decision, since software components (runnables) are given in compilable C-codes which require a selected RTOS code and run-time environment. Therefore, all necessary data, such as runnable size, dependency, their execution time, as well as label sizes and execution semantics are described by relatively simple AMALTHEA format. The possibility of optimizing the runnable and task mapping of an application described by an AMALTHEA file is presented later in this paper.

\section{AMALTHEA file format}\label{sec:FileFormat}

Despite being a rather compact format, AMALTHEA includes enough information for solving different system design problems. In this section, this format is briefly described.

Both hardware and software models of the system under development are stored in a file based on the XML format. The hardware model describes a system typically composed of a number of ECUs, microcontrollers, cores, memories, network, etc. Software model describes processes, labels and activations.

The processes generalize tasks and Interrupt Service Routines (ISR). They may be characterised with certain priority, used during process scheduling, or activation represented by certain stimulus. Any process can contain calls to other tasks or ISRs together with their execution orders. Alternative execution paths can be defined based on label values or probabilities. A process can activate another process, invoke a runnable or synchronise tasks using events issued by operating system. Runnables communicate with each other using
labels, data elements located in memory with certain bitlength. Activations may be single, sporadic, periodic or performed according to a custom pattern. Stimulus and clock objects are described in \emph{stimuli model}.

\section{DemoCar example}\label{sec:DemoCar}

We evaluate the AMALTHEA framework with a simplified model of an engine control application named DemoCar.  It consists of 6 tasks presented in Table \ref{tab:Tasks}. Four of them are periodic, activated respectively every 5ms, 10ms, 20ms, and 100ms, with relative deadlines equal to the periods. Two tasks are aperiodic: CylNumTriggeredTask and ActuatorTask, which are invoked by an RTOS. These tasks have no deadlines.

\begin{figure*}
\begin{center}
\includegraphics[width=0.8\textwidth]{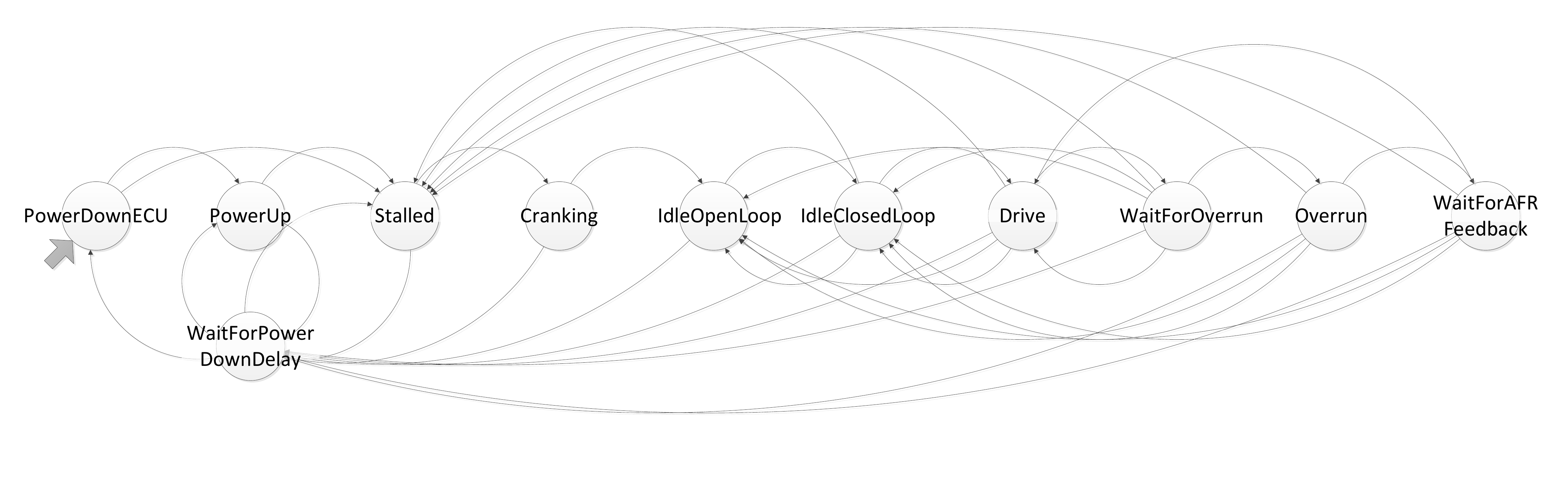}
\end{center}
\caption{DemoCar operating modes described by a Finite State Machine stored in OperatingModeSWCEntity runnable}\label{fig:FSM}
\end{figure*}

The runnables executed by these tasks are listed in Tab. \ref{tab:Runnables}. The labels used in the DemoCar use case are itemised in Tab. \ref{tab:Labels}. They are read and written by runnables, as indicated in the 3rd and 4th columns of Table \ref{tab:Runnables}. For example, runnable OperatingModeSWCEntity in Task20ms determines the mode in which the engine is currently operating using six read labels. These labels form conditions for mode changing, which is illustrated in Fig. \ref{fig:FSM}. To make this figure more readable, these conditions have been removed from the figure, but for example to change the current mode from PowerUp to Stalled, expression $PowerUpComplete \wedge IgnitionOn$ shall be true, whereas to move to WaitForPowerDownDelay from all the states drawn on the right side of PowerUp $IgnitionOn$ must equal false.

\begin{figure*}
\includegraphics[width=\textwidth]{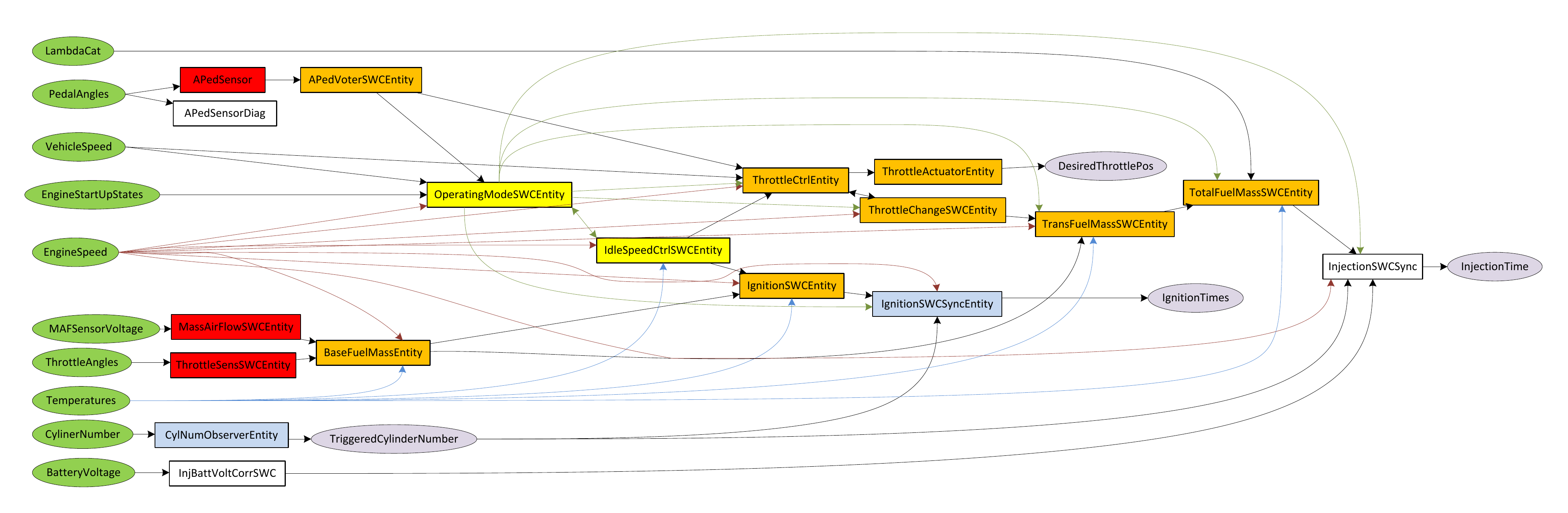}
\caption{Static structure of DemoCar ECU}\label{fig:StaticStructure}
\end{figure*}

The static structure of DemoCar is visualized in Figure \ref{fig:StaticStructure}. Labels are depicted as ellipses and Runnables as rectangles. The color for labels denotes the operation performed. Labels to be read are green, whereas labels to be written are light-violet. The color of runnables represents the period of their activation. Red stands for high activation rate (5ms period), orange for 10ms, yellow for 20ms and white for 100ms period. Light-blue rectangles indicate runnables started on an asynchronous event, either on message-reception or starting of the system. To improve the readability of this figure, some arrows have been drawn with different colors which have no particular meanings.

In the AMALTHEA format, each runnable contains an instruction list which includes a set of labels to be read and write to, and the runnable computation time. For example, in runnable APedSensor, two voltage values from two sensors (stored in labels APedSensor1Voltage and APedSensor2Voltage, reflecting the current accelerator pedal position, represented as PedalAngles in Figure \ref{fig:StaticStructure}) are read. Then, they are transformed into the corresponding percentage value and written to other labels: APedPosition1 and APedPosition2. EcuBrakePedalSensor reads the brake pedal position in voltage delivered by the sensor and translates it to a percentage of brake pedal position. MassAirFlowSensor reads the current mass air flow value from sensor in voltage and translates it to a value in kg/h. ThrottleSensor reads the throttle position from two redundant sensors in voltage and transforms them to percentage and determines a voted throttle position value. APedVoter reads the accelerator pedal positions and computes a voted accelerator pedal position. BaseFuelMass reads the mass air flow from MassAirFlowSensor and determines the base fuel mass per stroke. CylNumObserver observes requests for the injection time and ignition time parameters for specific cylinders. IgnitionTimeActuation reads the pre-calculated ignition time value from IgnitionTimingSWC. IgnitionTiming reads the determined mass air from BaseFuelMass and determines the optimal ignition time for combustion process. InjectionTimeActuation reads the pre-calculated total fuel mass per stroke from TotalFuelMass and transforms it to an injection time value. ThrottleActuator takes the new throttle position to be set as a percentage value and transforms it to a voltage value. ThrottleController determines the desired throttle position using the voted pedal position and determines the new throttle position to be set. TotalFuelMass reads the determined transient fuel mass per stroke from TransientFuelMass and determines the total fuel mass per stroke. TransientFuelMass  reads the base fuel mass per stroke from BaseFuelMass and compensates this mass for specific wall-wetting effects in the intake system.
More details on some of these runnables are provided in \cite{Frey10}.

It is worth noting the compactness of the format. For example, APedSensor is described by only 15 xml lines, whereas its total code in C language requires 163564 bytes. Despite this brevity, it is possible to extract all dependencies, as the labels to be read and written by this runnable are provided explicitly. Additionally, the memory footprint size, and execution time distribution is also provided.

\begin{table}
\caption{Tasks in DemoCar benchmark}\label{tab:Tasks}
\scriptsize{\begin{tabular}{p{0.3\columnwidth}|p{0.4\columnwidth}p{0.1\columnwidth}}
Name & Activation & Priority \\ \hline
CylNumTriggeredTask & aperiodic &  30\\
ActuatorTask & aperiodic &  25\\
Task5ms & period 5ms & 20\\
Task10ms & period 10ms & 15\\
Task20ms & period 20ms & 10\\
Task100ms & period 100ms & 5\\
\end{tabular}}
\end{table}

\begin{table*}
\caption{Runnables, their memory footprint in bits (Size), read and written labels, lower (Best Case Execution Time - BCET) and upper bound (Worst Case Execution Time - WCET) of execution time (in $\mu$s)}\label{tab:Runnables}
\scriptsize{\begin{tabular}{p{0.02\textwidth}|p{0.17\textwidth}|p{0.05\textwidth}p{0.27\textwidth}p{0.27\textwidth}p{0.04\textwidth}p{0.04\textwidth}}
Task & Runnable	&	Size 	&	Read labels &	Written labels	&	BCET	&	WCET\\ \hline
\parbox[t]{2mm}{\multirow{6}{*}{\rotatebox[origin=c]{90}{CylNum-}}}\parbox[t]{2mm}{\multirow{6}{*}{\rotatebox[origin=c]{90}{TriggeredTask}}} && \\
&& \\
& CylNumObserverEntity	&	55600	&	CylinderNumber	&	TriggeredCylinderNumber	&	434	&	1145\\
&& \\
&& \\
&& \\
\hline
\parbox[t]{2mm}{\multirow{6}{*}{\rotatebox[origin=c]{90}{Actuator-}}}\parbox[t]{2mm}{\multirow{6}{*}{\rotatebox[origin=c]{90}{Task}}}
&  IgnitionSWCSyncEntity	&	72512	&	IgnitionTiming, EngineSpeed, TriggeredCylinderNumber	&	IgnitionTime1, IgnitionTime2, IgnitionTime3, IgnitionTime4, IgnitionTime5, IgnitionTime6, IgnitionTime7, IgnitionTime8	&	2728	&	4921\\
&  InjectionSWCSync	&	69824	&	TotalFuelMassPerStroke, CrankFlag, TriggeredCylinderNumber, EngineSpeed, BatVoltCorr	&	InjTimeCyl1, InjTimeCyl2, InjTimeCyl3, InjTimeCyl4, InjTimeCyl5, InjTimeCyl6, InjTimeCyl7, InjTimeCyl8	&	1644	&	3302\\
\hline
\parbox[t]{5mm}{\multirow{4}{*}{\rotatebox[origin=c]{90}{Task5ms}}}
& MassAirFlowSWCEntity	&	56608	&	MAFSensorVoltage	&	MAFSensor	&	55	&	172\\
& ThrottleSensSWCEntity	&	58816	&	ThrottleAngle1, ThrottleAngle2	&	ThrottlePosition1, ThrottlePosition2	&	113	&	337\\
& APedSensor	&	66288	&	PedalAngle1, PedalAngle2	&	AcceleratorPedalPosition1, AcceleratorPedalPosition2	&	555	&	964\\ \hline
\parbox[t]{5mm}{\multirow{25}{*}{\rotatebox[origin=c]{90}{Task10ms}}}
& APedVoterSWCEntity	&	56832	&	AcceleratorPedalPosition1, AcceleratorPedalPosition2	&	VotedPedalPosition	&	87	&	287\\
& ThrottleCtrlEntity	&	70944	&	CoolantTemperature, EngineSpeed, MAFSensor, ThrottlePosition1, ThrottlePosition2	&	BaseFuelMassPerStroke, MafRateOut	&	3664	&	5783\\
& ThrottleActuatorEntity	&	128464	&	CoolantTemperature, CrankFlag, DesiredThrottlePosOut, EngineSpeed, FuelEnabled, InletAirTemperature, OverrunFlag, UpdatePeriod	&	RateOfThrottleChange, ThrottleImpulseBeta1, ThrottleImpulseBeta2	&	3788	&	5913\\
& BaseFuelMassEntity	&	70944	&	CoolantTemperature, EngineSpeed, MAFSensor, ThrottlePosition1, ThrottlePosition2 	&	BaseFuelMassPerStroke, MafRateOut	&	3664	&	5783\\
&
ThrottleChangeSWCEntity	&	128464	&	CoolantTemperature, CrankFlag, DesiredThrottlePosOut, EngineSpeed, FuelEnabled, InletAirTemperature, OverrunFlag, UpdatePeriod	&	RateOfThrottleChange, ThrottleImpulseBeta1, ThrottleImpulseBeta2	&	3788	&	5913\\
&
TransFuelMassSWCEntity	&	128464	&	InletAirTemperature, CoolantTemperature, MafRateOut, EngineSpeed, UpdatePeriod, RateOfThrottleChange, ThrottleImpulseBeta1, ThrottleImpulseBeta2, OverrunFuelShutoffFlag, CrankFlag, FuelEnabled, BaseFuelMassPerStroke	&	TransientFuelMassPerStroke	&	3985	&	6376\\
& IgnitionSWCEntity	&	66784	&	CrankFlag, MafRateOut, EngineSpeed, InletAirTemperature, OverrunIgnitionRetard, IdleFlag, IdleOLFlag, IdleIgnitionCorrection, CoolantTemperature	&	IgnitionTiming	&	3047	&	4537\\
& TotalFuelMassSWCEntity	&	66432	&	CrankFlag, LambdaCat1, LambdaCat2, CoolantTemperature, OverrunFuelShutoffFlag, TransientFuelMassPerStroke	&	TotalFuelMassPerStroke	&	743	&	1354\\ \hline
\parbox[t]{5mm}{\multirow{6}{*}{\rotatebox[origin=c]{90}{Task20ms}}}
& OperatingModeSWCEntity	&	139392	&	EngineSpeed, VehicleSpeed, IgnitionOn, PowerUpComplete, VotedPedalPosition, IdleSpeedSetpoint	&	OverrunFuelShutoffFlag, IdleFlag, IdleOLFlag, CrankFlag, OverrunFlag, FuelEnabled, AFRFeedbackFlag, OverrunIgnitionRetard, UpdatePeriod	&	18612	&	39281\\
& IdleSpeedCtrlSWCEntity	&	66976	&	IdleFlag, EngineSpeed, CoolantTemperature	&	IdleSpeedSetpoint, IdleThrottleCorrection, IdleIgnitionCorrection	&	913	&	1686\\
\hline
\parbox[t]{5mm}{\multirow{4}{*}{\rotatebox[origin=c]{90}{Task100ms}}}
&& \\
& APedSensorDiag	&	66288	&	PedalAngle1, PedalAngle2	&		&	102	&	235\\
& InjBattVoltCorrSWC	&	56928	&	BatteryVoltage	&	BatVoltCorr	&	290	&	547\\
&& \\

\end{tabular}}
\end{table*}

\begin{table}
\caption{Labels in DemoCar benchmark and their lengths (in bits)}\label{tab:Labels}
\scriptsize{
\begin{tabular}{l|l||l|l}
Name & Bitlength & Name & Bitlength\\ \hline
AcceleratorPedalPosition1	&	16	&	InjTimeCyl3	&	16\\
AcceleratorPedalPosition2	&	16	&	InjTimeCyl4	&	16\\
AcceleratorPedalPositions	&	16	&	InjTimeCyl5	&	16\\
AFRFeedbackFlag	&	1	&	InjTimeCyl6	&	16\\
BaseFuelMassPerStroke	&	16	&	InjTimeCyl7	&	16\\
BatteryVoltage	&	16	&	InjTimeCyl8	&	16\\
BatVoltCorr	&	16	&	InletAirTemperature	&	8\\
CoolantTemperature	&	8	&	LambdaCat1	&	16\\
CrankFlag	&	1	&	LambdaCat2	&	16\\
CylinderNumber	&	8	&	MafRateOut	&	16\\
DesiredThrottlePos	&	16	&	MAFSensor	&	16\\
DesiredThrottlePosOut	&	16	&	MAFSensorVoltage	&	8\\
EngineSpeed	&	16	&	OverrunFlag	&	1\\
FuelEnabled	&	1	&	OverrunFuelShutoffFlag	&	1\\
IdleFlag	&	1	&	OverrunIgnitionRetard	&	8\\
IdleIgnitionCorrection	&	8	&	PedalAngle1	&	16\\
IdleOLFlag	&	1	&	PedalAngle2	&	16\\
IdleSpeedSetpoint	&	16	&	PowerUpComplete	&	1\\
IdleThrottleCorrection	&	16	&	RateOfThrottleChange	&	16\\
IgnitionOn	&	1	&	ThrottleAngle1	&	16\\
IgnitionTime1	&	16	&	ThrottleAngle2	&	16\\
IgnitionTime2	&	16	&	ThrottleImpulseBeta1	&	16\\
IgnitionTime3	&	16	&	ThrottleImpulseBeta2	&	16\\
IgnitionTime4	&	16	&	ThrottlePosition1	&	16\\
IgnitionTime5	&	16	&	ThrottlePosition2	&	16\\
IgnitionTime6	&	16	&	TotalFuelMassPerStroke	&	16\\
IgnitionTime7	&	16	&	TransientFuelMassPerStroke	&	16\\
IgnitionTime8	&	16	&	TriggeredCylinderNumber	&	8\\
IgnitionTiming	&	8	&	UpdatePeriod	&	16\\
InjTimeCyl1	&	16	&	VehicleSpeed	    &	16\\
InjTimeCyl2	&	16	&	VotedPedalPosition	&	16\\
\end{tabular}}
\end{table}

\section{Parsing AMALTHEA model}\label{sec:Parser}

\begin{figure}
\includegraphics[width=0.5\textwidth]{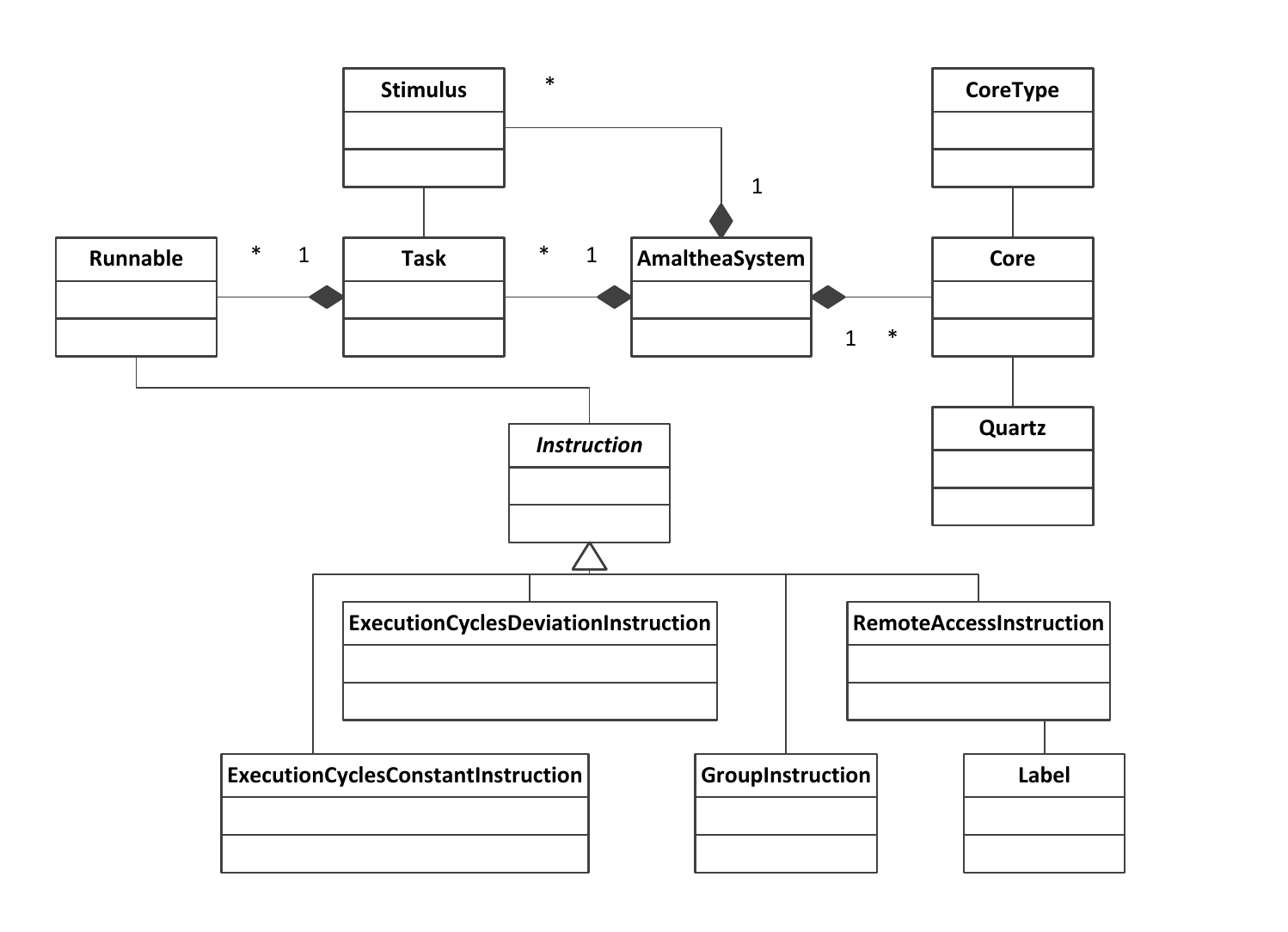}
\caption{UML class diagram of AMALTHEA file parser}\label{fig:ParserUML}
\end{figure}

One of the issues hindering the community from using AMALTHEA files as benchmark for research is the lack of C++ based parser in the platform release. To overcome this limitation, we have developed a software module to read and parse AMALTHEA file format. The main classes of this module, together with their dependencies, are presented in Figure \ref{fig:ParserUML}. In the proposed implementation, Apache Xerces C++\footnote{https://xerces.apache.org/xerces-c/}, a third-party tool for parsing, validating, serializing and manipulating XML, is used by the AmaltheaSystem singleton class. This class includes containers (both lists and maps) for main entities present in the AMALTHEA file together with the appropriate member access functions. It is treated as the entry point to the module whose interface can be used to fetch both the software and hardware models from the AMALTHEA file.

\subsection{Parsing AMALTHEA Software Model}
The classes representing entities of the AMALTHEA software model are depicted left and beneath the AmaltheaSystem class in Figure \ref{fig:ParserUML}. The Label class represents data elements, Runnable represents executable entities that perform calculation and read/write accesses to labels, Task represents clusters of runnables. The objects of these classes are created and stored by the AmaltheaSystem class. Its member function can be used to obtain the needed data. The most important functions returning information about the software model fetched from an AMALTHEA file are summarised in Table~\ref{tab:swModelFunctions}.

The AMALTHEA software model is stored in the following way. For each runnable, a job object is created, whereas the runnable execution order defined inside a task is reflected into dependencies of these job objects. All the dependencies described by stimuli are reproduced in a similar manner. 

\begin{table}
\caption{AmaltheaSystem class member functions returning information fetched from an AMALTHEA file software model}\label{tab:swModelFunctions}
\begin{center}
\scriptsize{\begin{tabular}{p{0.3\columnwidth}|p{0.6\columnwidth}}
Function	&  Description \\ \hline
Label* GetLabel(int Index);	& Returns Label object stored in the internal list in the Index position\\
int GetNoOfLabels();	& Returns the number of Label objects stored in the internal list\\
Label* GetLabel(string IDIn);	& Returns Label object with the provided ID\\
Label* GetLabelByName(string NameIn);	& Returns Label object with the provided name\\
int GetIndexOfLabel(Label* LabelIn);	& Returns position of the provided Label object in the internal list\\
Runnable* GetRunnable(int Index);	& Returns Runnable object stored in the internal list in the Index position\\
Runnable* GetRunnable(string IDIn);	& Returns Runnable object with the provided ID\\
int GetNoOfRunnables();	& Returns the number of Runnable objects stored in the internal list\\
int GetIndexOfRunnable(Runnable* RunnableIn);	& Returns position of the provided Runnable object in the internal list\\
Stimulus* GetStimulus(int Index);	& Returns Stimulus object stored in the internal list in the Index position\\
int GetNoOfStimuli();	& Returns the number of Stimulus objects stored in the internal list\\
Task* GetTask(int Index);	& Returns Task object stored in the internal list in the Index position\\
int GetNoOfTasks();	& Returns the number of Task objects stored in the internal list\\
Label* GetLabelWithIndex(int Index);	& Returns Label object stored in the internal list in the Index position\\
Runnable* GetLabelsDest(Label* LabelIn);	& Returns a Runnable object reading from the provided Label object\\
Runnable* GetLabelsSource(Label* LabelIn);	& Returns a Runnable object writing to the provided Label object\\
\end{tabular}}
\end{center}
\end{table}

\subsection{Parsing AMALTHEA Hardware Model}
The classes representing entities of the AMALTHEA hardware model are depicted right to the AmaltheaSystem class in Figure \ref{fig:ParserUML}. The Core class represents a processing core executing tasks. Each core is associated with a particular CoreType element, where the number of clock ticks needed for executing a single instruction is provided. Each core is connected with a Quartz object, providing a clock frequency. In the current version of implementation, we ignore the information about memories stored in the AMALTHEA file, but we plan to add it in future releases of the parsing module. The member functions returning information about the hardware model fetched from an AMALTHEA file are summarised in Table~\ref{tab:hwModelFunctions}.

\begin{table}
\caption{Amalthea System class member functions returning information fetched from an AMALTHEA file hardware model}\label{tab:hwModelFunctions}
\begin{center}
\scriptsize{\begin{tabular}{p{0.4\columnwidth}|p{0.5\columnwidth}}
Function	&  Description \\ \hline
CoreType* GetCoreType(string IDIn);	& Returns CoreType object with the provided ID \\
Quartz* GetQuartz(string IDIn);	& Returns Quartz object with the provided ID \\
Core* GetCore(string IDIn);	& Returns Core object with the provided ID \\
Core* GetCore(int Index);	& Returns Core object stored in the internal list in the Index position \\
int GetNoOfCores();	& Returns the number of Core objects stored in the internal list \\
\end{tabular}}
\end{center}
\end{table}

\section{Case study: resource allocation for Automotive Democar benchmark represented in AMALTHEA format}\label{sec:CaseStudy}

\begin{figure}
\includegraphics[width=0.5\textwidth]{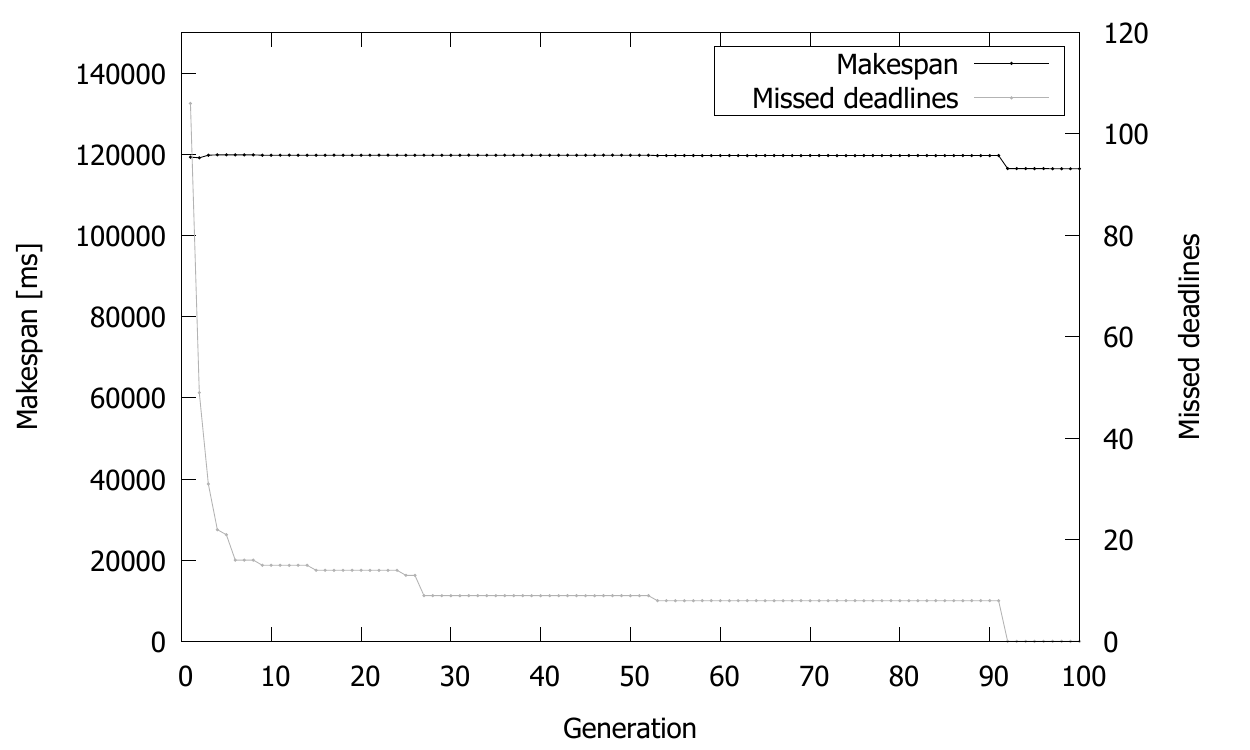}
\includegraphics[width=0.5\textwidth]{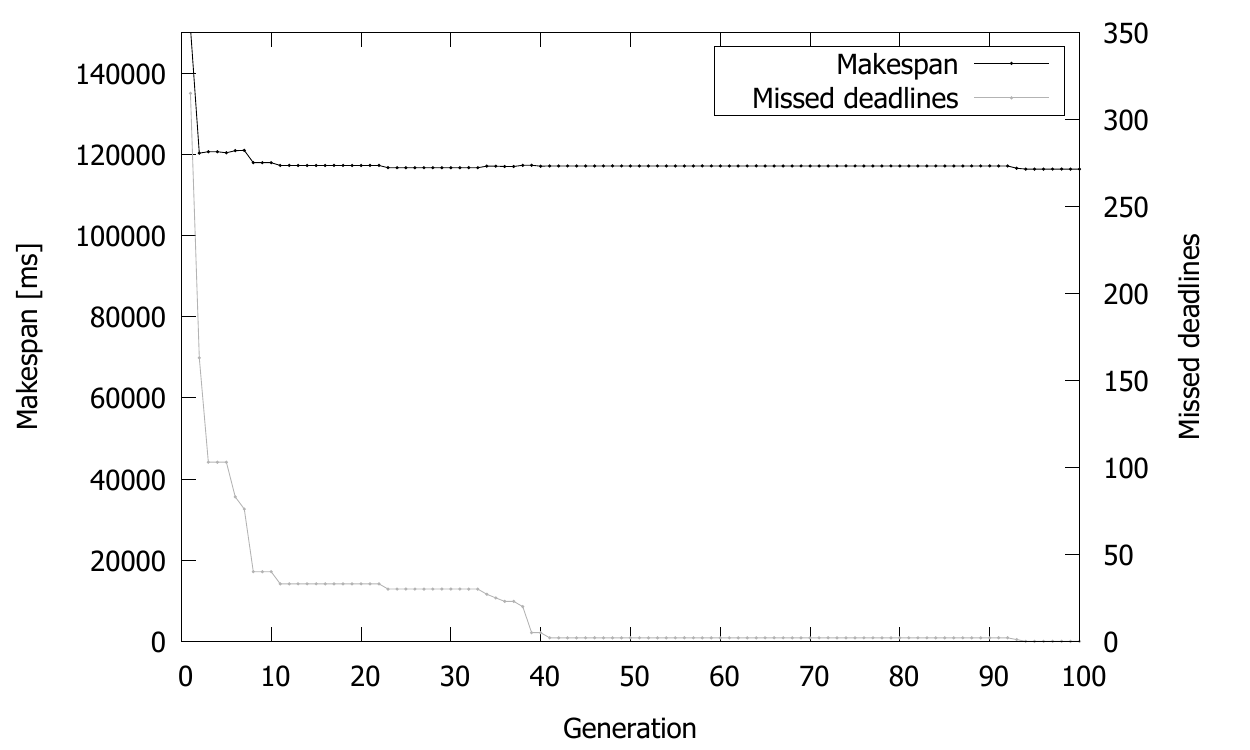}

\caption{Missed deadlines (grey) and makespan (black) value optimization for DemoCar implemented on 2x2 mesh-based NoC with four (above) and three (below) active cores}\label{fig:DemoCarNoCTwoSizes}
\end{figure}

For the example DemoCar benchmark to be executed on a multi-core embedded system, we evaluate makespan (also known as response time) and number of violated deadlines during one hyperperiod (i.e., the least common multiple of all runnables' periods) by allocating runnables and labels to different cores. In this experiment, we use a genetic algorithm that aims to explore the allocation space towards achieving solutions with optimised timing behaviour \cite{Sayuti13}. For demonstration, a Network on Chip (NoC) based multi-core platform with XY routing algorithm has been chosen.  For the DemoCar application, the size of the mesh has been initially configured as 2x2. The application model has been extended with communication messages between tasks and labels. The genetic algorithm is then executed to perform both task and label allocations to cores during 100 generations of 20 individuals each. The makespan has been evaluated using the technique described in \cite{SAMOS}. The first fully schedulable allocation has been found in the 92-nd generation. The makespan value decreases with the generation number, which shows that optimized allocations are achieved with increased generation
 (Figure \ref{fig:DemoCarNoCTwoSizes} top).

Then one core of the NoC has been switched off for energy conservation reason. For this 3-core platform, the fully schedulable allocation has been found in the 94-th generation (Figure \ref{fig:DemoCarNoCTwoSizes} bottom).  
A fully schedulable allocation has not been found for 2x2 mesh NoC with two disabled cores, despite analysing much wider search space than previously - spanning over four islands with 100 individuals each. The best found allocation leads to violation of 216 out of 1204 deadlines.  The average execution time of performing the DemoCar schedulability analysis has been less than 0.1s, whereas the total parsing process lasted 0.008s (the computation has been performed on one core of a typical desktop computer).

\section{Conclusions}\label{sec:Conclusions}

AMALTHEA open source development environment for multi-core systems can be used to generate automotive benchmarks for real-time community striving for close-to-life use cases, similar to demonstrated DemoCar example. This system has been described in details so that it can be reconstructed in other file formats.

To facilitate using AMALTHEA to generate benchmarks, an AMALTHEA format parser has been developed and published. This parser has been integrated with a genetic algorithm framework and schedulability analysis to perform a functionality to multi-core system resources allocation optimisation.

\end{document}